\documentclass[useAMS,usenatbib]{mn2e}

\usepackage{graphicx}
\newcommand{\gae}{\lower 2pt \hbox{$\, \buildrel {\scriptstyle >}\over {\scriptstyle\sim}\,$}}
\newcommand{\lae}{\lower 2pt \hbox{$\, \buildrel {\scriptstyle <}\over {\scriptstyle\sim}\,$}}
\usepackage[compact]{titlesec}

\title[]{ 0.5 Mpc-scale extended X-ray emission in 4C 23.56}
\author[O. Johnson et al.]{O. Johnson$^{1}$\thanks{E-mail:
cocj@roe.ac.uk (OJ)}, O. Almaini$^{2}$, P. N. Best$^{1}$, J. Dunlop$^{1}$\\
$^1$SUPA\thanks{Scottish Universities Physics Alliance}, University of Edinburgh, Institute for Astronomy, Royal Observatory, Edinburgh EH9 3HJ, UK\\
$^{2}$School of Physics \& Astronomy, University of Nottingham, University Park, Nottingham NG7 2RD, UK}
\begin{document}

\date{}


\maketitle

\begin{abstract}
We present an XMM-Newton observation of the radio galaxy 4C 23.56 at $z=2.48$ which reveals extended X-ray emission coincident with the radio lobes spanning $\sim 0.5$ Mpc.  These are the largest X-ray-bright lobes known at $z>2$.  Under the assumption that these features result from inverse-Compton scattering of cosmic microwave background photons by relativistic electrons in the radio source lobes, the measured luminosity of $L_{\mathrm{0.5-8 keV}}=7.5\times10^{44}$ erg s$^{-1}$ implies a minimum energy stored within the lobes of $\sim10^{59}$ erg.  We discuss the potential of the large-scale radio/X-ray lobes to influence evolution of the host galaxy and proto-cluster environment at high redshift. 
\end{abstract}


\section{Introduction}

The galaxies hosting powerful radio-loud active galactic nuclei (AGN) at high
redshift are believed to evolve into the most massive galaxies in the
present-day Universe, which reside locally in rich cluster environments (e.g., Best et al., 1998).
High-redshift radio galaxies (HzRGs) are therefore popular targets
in studies of the co-evolution of massive galaxies, their central AGN, and
their environment on larger scales.

In recent years, observations with Chandra and XMM-Newton have detected
extended X-ray emission in an increasing sample of HzRGs 
(Fabian et al. 2003, 2003b; Carilli et al. 2002; Yuan et al. 2003; Scharf et al. 2003; 
Belsole et al. 2004; Overzier et al. 2005; Erlund et al. 2006;
Blundell et al. 2006).  Generally associated with the radio
structure, the extended X-ray emission is believed to arise in most cases partly or wholly
from inverse-Compton (IC) scattering of cosmic microwave background (CMB)
photons by relativistic electrons in the jets and/or lobes of the radio
galaxy.  As noted by Schwartz (2002), the steep $(1+z)^4$ dependence of the
density of CMB photons on redshift suggests this process will be increasingly
significant toward higher redshifts.  Based on extrapolation from the radio
source number counts, Celotti \& Fabian (2004) estimate that at redshifts
above $z\gae1$ the majority of luminous ($L_X>10^{44}$ erg s$^{-1}$) extended
X-ray sources may result from IC scattering of the CMB in HzRGs.

This new class of X-ray source may provide crucial insight into the effect of
AGN activity on the evolution of massive galaxies and their larger scale
environment.  The identification of bubble-like features in low-and
intermediate-redshift clusters indicates AGN jets can inject a significant
amount of energy into their environs (e.g., McNamara et al. 2000; Fabian et
al. 2006).  Similar processes have been suspected to exert significant
influence on the evolution of structure at high redshift, e.g., through the
regulation of gas accretion and star formation in massive galaxies (e.g., Silk \& Rees, 1998)
and as a source of non-gravitational heating within forming 
cluster cores (e.g., Bower et al., 2001).
However, the mechanisms and magnitude of AGN feedback at high redshift are
unknown.

\begin{figure*}
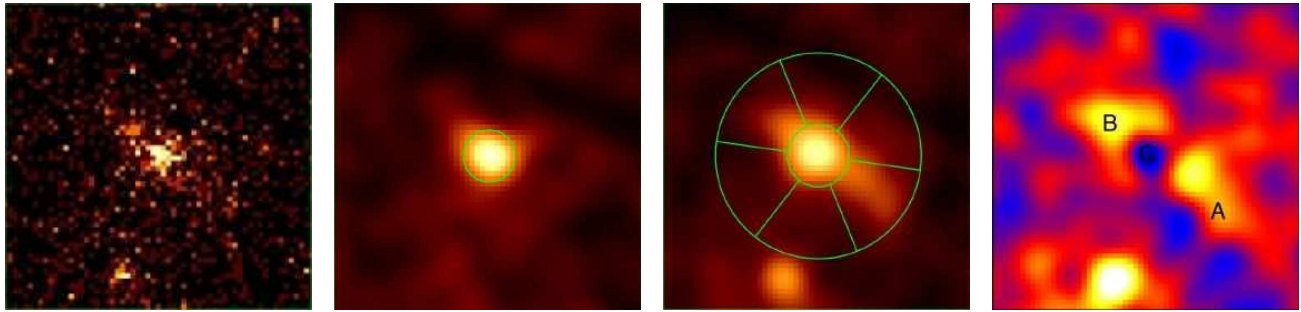

\begin{minipage}{\textwidth}
\caption{{\it a:} 0.5--8 keV EPIC image of 4C 23.56.  \label{fig:image_raw} {\it b:} Smoothed 2--8 keV EPIC image of 4C 23.56, with the region used for spectral analysis of the X-ray point source overlaid.  The darker line seen through the NW corner is a chip gap in the pn camera. \label{fig:image_soft}{\it c:} Smoothed 0.5--2 keV EPIC image of 4C 23.56, with regions used for spectral analysis of the extended emission overlaid.  {\it d:} Difference image obtained by subtracting a smoothed 0.5--2 keV band image from a smoothed 2--8 keV band image.  Soft flux appears light (yellow) and hard flux appears dark (blue).  Labels indicate the Knopp \& Chambers (1997) 6 cm radio positions of the 4C 23.56 core (C) and NE (B) and SW (A) lobes.  \label{fig:image_sub}  Images are $2\arcmin$ on a side and orientated so that North is up and East is to the left.}
\begin{center}
\includegraphics[width=0.23\textwidth, angle=0]{epic_raw.epsf}
\hspace{1mm}
\includegraphics[width=0.23\textwidth, angle=0]{pn_circle_colour.epsf}
\hspace{1mm}
\includegraphics[width=0.23\textwidth, angle=0]{pn_panda_colour.epsf}
\hspace{1mm}
\includegraphics[width=0.23\textwidth, angle=0]{ep_hmins_colour3.epsf}
\end{center}
\end{minipage}
\end{figure*} 

The detection of IC-scattered X-ray emission in HzRGs has begun to clarify the
evolution and energetics of high-redshift radio lobes.  As energy losses from
the more energetic electron population responsible for the radio emission are
more rapid than those of the less energetic population involved in IC
scattering, the X-ray lobe emission can outlast the radio emission and provide
a measure of past radio activity (e.g., Fabian et al. 2003; Blundell et
al. 2006).  X-ray measurements have allowed estimates of the minimum Lorentz
factors, $\gamma_{min}$, in radio hotspots (Blundell et al., 2006) and the
total energy contained within X-ray luminous radio lobes (Erlund et al. 2006)
at $z\sim2$.  Scharf et al. (2003), noting the relative positions of X-ray and
Ly$\alpha$ flux in 4C 41.17 at $z=3.8$, suggest that the IC-scattered X-ray
photons could themselves give rise to a secondary feedback mechanism by
photoionising the surrounding medium.  Both the direct mechanical feedback
from the jets and this second radiative mechanism could influence the
structure and energetics of forming structures on galactic and cluster scales
and, interestingly, would preferentially affect the most massive structures at
high redshift.

In this Letter we report the detection of extended X-ray emission associated with the radio galaxy 4C 23.56  ($z=2.483$; R{\"o}ttgering et al., 1997) in XMM-Newton observations obtained as part of a larger study of the X-ray properties of HzRGs. We begin with a summary of the multiwavelength properties of 4C 23.56 in \S2, present the X-ray observations and analysis in \S3 and briefly discuss our findings in \S4.  Throughout this letter we assume a $\Lambda$CDM cosmology with $\Omega_m$=0.3, $\Omega_{\Lambda}=0.7$, and $H_0=70$ km s$^{-1}$ kpc$^{-2}$.  Unless specified otherwise, X-ray fluxes are given in the observed-frame 0.5--8 keV band, X-ray luminosities are corrected for Galactic absorption and given in the rest-frame 0.5--8 keV band, and errors are quoted at the 90\% confidence level.        

\section{4C 23.56}
4C 23.56 is an ultra-steep spectrum, Fanaroff-Riley class II radio galaxy which is among the largest known at high redshift.  The core and radio lobes lie along an axis with a position angle (P.A.) of $52^\circ$ E of N and an angular size of roughly $0.5$ Mpc (Chambers et al., 1996a).  Multiwavelength observations indicate a heavily obscured AGN in a dusty host galaxy (Chambers et al., 1996b; Knopp \& Chambers, 1997, Archibald et al., 2001).  An enormous Ly-$\alpha$ emission-line nebula comprised of two opposing ionization cones with opening angles of $\sim90^{\circ}$ covers an area of $\sim 5000$ kpc$^2$ parallel to the radio axis.  The near-symmetry of the emission-line morphology indicates the radio axis lies nearly in the plane of the sky, while the relative brightness of the southwestern lobe implies it is somewhat nearer to us.  Moderate excess counts of both Ly$\alpha$ emitters and faint, red galaxies in the vicinity of 4C 23.56 suggest it may reside in an over-dense cluster or proto-cluster environment (Knopp \& Chambers, 1997).   

\section{X-ray observations}
\label{observations}

We obtained 32 ks XMM-Newton observations of the 4C 23.56 field centred at 21$^{h}$07$^{m}$12.70$^{s}$, 23$^{\circ}$31$^{'}$23.0$^{''}$ on 29 November 2004 (ObsID 0204400201).  The data were taken in Full Frame mode through the Thin filter and reduced using version 6.0.0 of the XMM Science Analysis System (SAS).  
The observations were minimally affected by periods of background flaring, which we identified in 7 -- 15 keV lightcurves and removed.  The remaining useful exposures were 27.7 ks in the pn camera and 30.8 ks in the MOS cameras.   We applied the standard recommended PATTERN, FLAG, and energy filters to the data to produce event lists suitable for imaging and spectral analysis.

\subsection{Imaging analysis}
\label{imaging}

We produced images for each of the EPIC cameras in the full (0.5--8 keV), soft (0.5--2 keV), and hard (2--8 keV) bands by binning the filtered events into $2\arcsec$ pixels.  We constructed a combined EPIC image by summing the individual instrument images.  To improve visibility, we also created smoothed versions of all images by convolving the counts images with a Gaussian kernel roughly the size of the XMM PSF ($\sigma=6\arcsec$).

Figure 1a shows a $2\arcmin\times2\arcmin$ region of the unsmoothed, full-band, EPIC image centred on the position of 4C 23.56.  The radio galaxy is clearly detected in the X-ray and appears slightly extended toward the NE and SW.  The extended emission is more easily identified in the smoothed images of the same region.  While the hard-band image (Figure 1b) shows only a point-source at the position of the radio galaxy, the soft-band emission (Figure 1c) is unambiguously extended along the NE-SW axis.  An unrelated soft X-ray point source, visible in the unsmoothed counts image, is also seen at the southern edge of the Figure.  

To quantify the significance of the extended emission detection, we extracted hard and soft band counts within the pie shaped regions centred on the X-ray point source shown in Figure 1c.  Two of the sectors are aligned with the axis of the extended emission while the other four serve as background regions.  To the extent that the XMM PSF is spherically symmetric, these regions should contain roughly the same amount of contamination from the central point source.  Across all three detectors, we find 144 soft-band counts within the two source regions and 156 counts within the four background regions.  Correcting for area, we expect $86$ background counts within the source region, implying an excess detection within the source regions of 58 counts, which is significant at the $8.6\sigma$ level.  The net source counts are split roughly evenly between the NE and SW regions.  Considering the pn and the MOS counts separately, we find significant soft-band detections of the extended flux at the $7.5\sigma$ and $4.8\sigma$ levels, respectively.  We find no statistically significant excess of hard-band counts.   

\begin{figure}
\caption{{\it Top:} X-ray spectra of the core of 4C 23.56 from the pn (upper points, black) and MOS (lower points, green and red) cameras, with the best-fitting model (histograms) described in the text overlaid.  {\it Bottom:} Fit residuals.
\label{fig:core_spec}}
\begin{center}
\includegraphics[width=0.44\textwidth, angle=0]{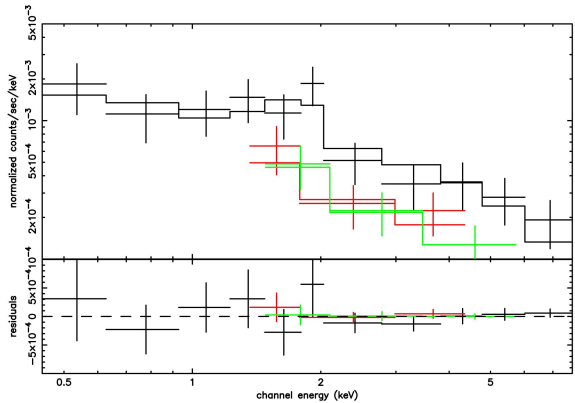}
\end{center}
\end{figure} 

To maximise the contrast of the faint, soft, extended flux, we subtracted the smoothed soft-band image from the smoothed hard-band image to produce the difference image shown in Figure 1d.  As both the central point source and the background have harder X-ray spectra than the extended emission, this method of displaying the data effectively de-emphasizes these sources of contamination, allowing the two lobes of soft emission to stand out distinctly on either side of the X-ray core.  We stress that the morphology of this difference image is not directly indicative of the surface brightness of the extended emission and must be interpreted with caution.  For example, the deficit of soft emission around the point source does not indicate the extended component is fainter near to the radio galaxy, rather that the contribution of hard flux in the wings of the point source is dominant over the soft emission at these radii.   

We indicate in Figure 1d the 6 cm radio positions of the core of 4C 23.56 and the peaks of its NE and SW radio lobes (Knopp \& Chambers, 1997).  The peak of the X-ray point source is coincident with the radio position of the core and the bulk of the extended X-ray emission is aligned with the radio lobes at P.A.$\sim52^{\circ}$.  The peak of the excess soft flux coincides with the NE radio lobe at a distance of $\sim20\arcsec$ from the radio core ($\sim160$ kpc at $z=2.483$).  In the SW, the excess soft flux appears to peak interior to the radio lobe, at a radius of $\sim20\arcsec$, and slightly out of alignment, forming an axis with the core with P.A.$\sim 68^{\circ}$.  A further, fainter extension of the excess soft X-ray flux just reaches the position of the SW radio lobe at a distance of $\sim40\arcsec$ ($\sim330$ kpc) from the centre.  The full extent of the extended emission along P.A.$\sim 52^{\circ}$ is $\sim70\arcsec$ ($\sim570$ kpc).  Unfortunately, the large XMM PSF and the substantial contamination of the soft emission by the wings of the central point source prohibit more precise characterisation of the morphology of the surface brightness of the extended emission.  

\subsection{Spectral analysis}

\begin{figure}
\caption{{\it Top:} X-ray spectra of the 4C 23.56 extended emission from the pn (upper points, black) and MOS (lower points, green and red) cameras, with the best-fitting model (histograms) described in the text overlaid.  {\it Bottom:} Fit residuals.
\label{fig:ext_spec}}
\begin{center}
\includegraphics[width=0.44\textwidth, angle=0]{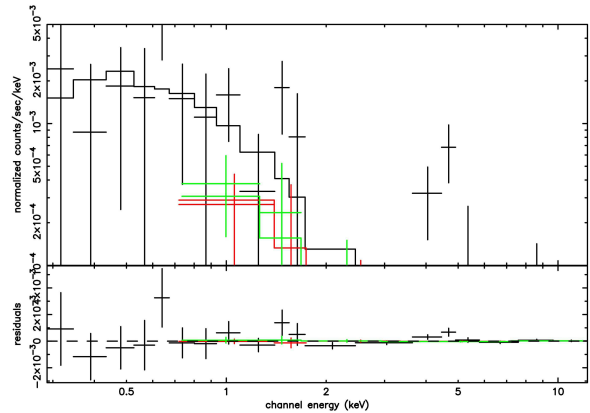}
\end{center}
\end{figure} 

We extracted X-ray spectra for the central point source and for the extended emission from each EPIC camera separately.  For the core emission, we chose a circular aperture of $10\arcsec$ radius shown in Figure 1b.  As the extended emission prohibited extraction of an annular background, suitable nearby regions were selected.  For the extended emission, we extracted source and background counts from the pie-shaped sectors described above and shown in Figure 1c.  Response files were created for each camera using the standard SAS {\tt especget} tool, which we ran in extended source mode for the lobe emission.  The spectra were grouped such that each bin contained 10 counts and spectra from the three cameras were fit simultaneously to models within the XSPEC package.  Fits were minimized using the $\chi$ Gehrels statistic as appropriate for low count data.  90\% confidence levels were estimated from projection of the statistic surface.  A Galactic absorption column of $N_H=1.28\times10^{21}$ cm$^{-3}$ was included in all the fits.

A single power-law fit indicates that the unresolved radio galaxy core is a hard X-ray source with a photon index $\Gamma_{core} = 1.0^{+0.3}_{-0.3}$.  Although we cannot distinguish between absorption and an intrinsically hard spectrum with the present data, such a hard spectrum usually indicates the presence of photoelectric absorption in an AGN.  Fixing the photon index to a typical $\Gamma_{core}=1.9$ (Miyaji et al., 2006), the best-fitting model (reduced $\chi^2$=1.2) yields a substantial column density of $N_{H}=1.4^{+1.8}_{-0.8}\times10^{23}$ cm$^{-2}$ and a luminosity for the underlying power-law of $1.8^{+0.8}_{-0.5}\times10^{45}$ erg s$^{-1}$, implying an aperture corrected luminosity of $L_{core}=3.6^{+1.5}_{-1.0}\times10^{45}$ erg s$^{-1}$.  The relatively poor fit statistic for the absorbed powerlaw results from the inability of this model to fit an excess of soft emission in the lowest-energy bins.  We therefore add a second powerlaw without local obscuration to represent the soft, extended component within the core aperture.  A much improved fit, shown in Figure \ref{fig:core_spec}, is obtained for a model with $\Gamma_{ext}=2.6$.  The complexity of the model strains the statistical limits of the dataset, however, and the four free parameters cannot be constrained simultaneously.  Freezing $\Gamma_{ext}$ at the preferred value of 2.6 yields an absorption of $N_{H}=5.2^{+3.2}_{-2.5}\times10^{23}$ cm$^{-2}$, a core luminosity of $L_{core}=3.6\times10^{45}$ erg s$^{-1}$, and a luminosity in the soft power-law component of $L_{ext}=7.3\times10^{44}$ erg s$^{-1}$.  However, the fit can only rigourously provide a joint upper limit to these luminosities of $4.6\times10^{45}$ erg s$^{-1}$.

The statistical quality of the spectra of the extended source is extremely
poor due to the faintness of the source emission.  As can be
seen in Figure \ref{fig:ext_spec}, there are only a few bins in the MOS
spectra and many of the bins in the pn spectrum are statistically equivalent
to zero.  The positive pn bins at 4--5 keV most likely result from imperfect
subtraction of the background.  We fit a single powerlaw model to the data.  
The best-fitting solution (reduced $\chi^2=0.83$)
yields a soft photon index $\Gamma_{e}=3.2^{+1.0}_{-0.9}$ and a
luminosity of $7.6\times10^{44}$ erg s$^{-1}$.  
Exclusion of the possibly contaminated bins at 4--5 keV does not
significantly alter the fit.  
For completeness, we note that an equally good fit (reduced $\chi^2=0.82$) was obtained
for a thermal Mekal model with a temperature of 1.7 keV, a
luminosity of $4.0\times10^{44}$ erg s$^{-1}$, and an abundance of 0.3 solar.  
The $3\sigma$ limit on the luminosity of any thermal emission on the scale 
of the detected lobes outside the radio 
axis is $2.1\times10^{44}$ erg s$^{-1}$.

We can also estimate the luminosity of the extended component directly from
the counts extracted from the pn camera in \S\ref{imaging}. 
 Following the procedure described in XMM Survey Science
Centre memo SSC-LUX\_TN-0059, Issue 3.0 and adopting a $\Gamma=3$ powerlaw spectrum with Galactic absorption, we find a 0.5--2 keV pn Energy Conversion Factor of
$6.804\times10^{11}$ counts cm$^{2}$ erg$^{-1}$.  Dividing the net counts in
the soft band by this value and the exposure time yields a soft-band flux of
$2.6\times10^{-15}$ erg s$^{-1}$ cm$^{-2}$, implying an unabsorbed full-band
luminosity of $7.5\pm2.0\times10^{44}$ erg s$^{-1}$ which is in excellent
agreement with the spectroscopically derived value.  

\section{Discussion}
\subsection{The nature of the extended X-ray emission}

The analyses above indicate soft, extended, X-ray emission coincident with the radio lobes of 4C 23.56 over a scale of $\sim0.5$ Mpc with a luminosity of  $\sim7\times10^{44}$ erg s$^{-1}$.  While there are several possible emission mechanisms for extended X-ray flux in HzRGs (see, e.g., Overzier et al., 2005 and references therein), few can account for the morphology and magnitude of the flux seen in 4C 23.56.  

Extrapolating the synchrotron spectrum from the radio waveband, we find the observed X-ray luminosity is two orders of magnitude higher than would be expected from synchrotron emission.  Synchrotron self-Compton emission, often invoked to explain X-ray hotspots in local radio galaxies (e.g., Harris et al. 2000), cannot produce X-rays over a large area because the synchrotron photon density can only exceed that of the CMB in compact regions.  IC scattering of optical/IR photons from the central AGN (Brunetti et al., 2001) can not account for emission so far from the core.  Shock heating of an ambient medium by the expanding radio source --- a proposed mechanism for the extended X-ray emission in PKS1138-262 (Carilli et al., 2002) --- would require an enormous reservoir of hot, high filling-factor gas which is not detected in the off-axis regions.  Finally, thermal emission from hot intracluster gas cannot be ruled out, but the surface brightness of the observed emission is much higher than expected and a contrived containment model would be needed to explain an extremely elongated morphology aligned with the radio lobes.   

We believe that in 4C 23.56, as in most previously reported HzRGs with X-ray-bright radio lobes, IC scattering of CMB photons is the most likely emission mechanism for the majority, if not the entirety, of the extended X-ray flux.  Fourteen $z>1$ HzRGs currently in the literature 
(see references in \S1) have X-ray lobes on scales ranging from a few tens to a few hundreds of kpc with typical luminosities of a few times $10^{44}$ erg s$^{-1}$.  The X-ray lobes in 4C 23.56 are the second largest discovered to date\footnote{The giant radio galaxy 6C 0905+3955 at $z=1.88$ exhibits X-ray emission associated with hotspots and lobes which are separated by $\sim 865$ kpc (Blundell et al. 2006).}, and the first $\gae 0.5$ Mpc-scale lobes discovered at the $z>2$ epoch suspected to be the predominant era of massive galaxy formation.

\subsection{Energetics of the X-ray-detected lobes}

We now briefly consider the energetics of the X-ray detected lobes, assuming the entirety of the extended X-ray emission arises in IC scattering of CMB photons.  Particles with a large Lorentz factor, $\gamma_{e}$, will upscatter CMB photons of frequency $\nu_{CMB}$ to higher frequencies $\nu_{scat}$ as

\begin{equation}
{\nu_{scat} \over \nu_{CMB}} \simeq {4 \over 3} \gamma_{e}^2.
\end{equation}

\noindent Scattering to X-ray energies therefore requires electrons with a Lorentz factor $\gamma_{e}\sim 10^3$.  If we assume that all electrons responsible for the X-ray lobes have this Lorentz factor, we can use the observed X-ray luminosity and the known CMB photon energy density to estimate the energy stored in the IC-emitting electron population.  The energy flux in scattered X-rays is

\begin{equation}
L_{X}=N_{e} U_{rad} {4 \over 3} \gamma_{e}^2 \sigma_{T} c,
\end{equation}

\noindent while the energy of the relativistic electrons is

\begin{equation}
E_{e}=N_{e} \gamma_{e} m_{e} c^2,
\end{equation}

\noindent where $N_{e}$ is the number density of relativistic electrons, $\sigma_{T}$ is the Thompson cross-section, $c$ is the speed of light, and $m_{e}$ is the rest mass of the electron.  $U_{rad}$ is the energy density of the CMB, which increases from the local value with redshift as $(1+z)^4$.  The energy in the scattering electron population is then given by

\begin{equation}
E_{e}={3 \over 4} {L_{X} m_{e} c \over U_{rad} \gamma_{e} \sigma_{T}} \simeq {3 \over 4} {L_{44} \over \gamma_{e}(1+z)^4} 10^{64}\, \mathrm{ erg,}
\end{equation}

\noindent where $L_{44}$ is the 2--10 keV X-ray luminosity in units of $10^{44}$ erg s$^{-1}$.

As noted by Erlund et al. (2006), this calculation provides only a lower limit on the energy in the scattering population.  Assuming a more realistic power-law distribution of Lorentz factors reveals dependencies of the total electron energy on the minimum and maximum Lorentz factors and on the X-ray photon index.  These parameters are poorly constrained; adopting typical values increases the estimated energy by a factor of a few.  In addition, this estimate considers only the energy in relativistic electrons.  A substantial population of relativistic protons (e.g., Celotti \& Fabian, 1993; Sikora et al., 1995) could further increase the implied lobe energies by a factor of up to $10^3$.  

Converting our measured 0.5--8 keV luminosity to 2--10 keV with $\Gamma=3$, we find a lower limit to the energy stored in relativistic electrons in the lobes of 4C 23.56 of $E_{e}\simeq8.0\times10^{58}$ erg; adopting a more typical $\Gamma=2$, $E_{e}\simeq2.2\times10^{59}$ erg.  Comparing these estimates with those of Erlund et al. (2006) for three $z\sim2$ HzRGs we note that while the X-ray lobes in 4C 23.56 are at least twice as large as the lower-redshift sources, the minimum energy stored in IC-emitting electrons appears to be roughly comparable.  Considering only losses through IC emission, the radiative lifetime of the 4C 23.56 X-ray lobes is $\sim10^{7}$ yr.

\subsection{Potential for environmental feedback}

An enormous reservoir of energy --- equivalent to $\sim 10^8$ supernovae --- is stored in the IC-emitting electron population within the radio lobes of 4C 23.56.  
As the cooling timescale of this population through IC-scattering to X-ray wavelengths is relatively long, a large fraction of this energy could potentially be injected into the surroundings during a given epoch of AGN activity. 
The vast extent of the X-ray luminous lobes suggests a mechanism by which significant feedback could take place at radii of up to 250 kpc from the central nucleus.  It is not clear, however, if the energy available in the IC scattering electron population is sufficient to be a major source of feedback influencing the evolution of either the host galaxy or the forming cluster environment.

Assuming a velocity dispersion of 300 km s$^{-1}$, the lobe energy estimates derived above are 10--20 times lower than the binding energy of a $10^{12}$ M$_{\odot}$ elliptical and roughly 500 times lower than
the $5\times10^{49}$ erg M$_{\odot}^{-1}$ estimated by Benson et al. (2003) to be required to inhibit the excessive growth of such a galaxy at high redshift.  In terms of heating of the intracluster medium, Wu, Fabian, \& Nelson (2000) estimate non-gravitational heating of 1--2 keV per particle is needed to explain the observed deviation of galaxy cluster properties from the theoretical scaling relations.  For a cluster with a gas mass of $10^{13}$ M$_{\odot}$, this is equivalent to $\sim5\times10^{61}$ erg or roughly 300 times our minimum energy estimate for the IC-scattering population in the lobes of 4C 23.56.  

Unless the assumption of a single Lorentz factor, $\gamma_{e}=10^3$, is dramatically wrong or there is a significant proton component to the jet, the energy contained in the IC-scattering population in the lobes appears to fall short of the required feedback energies for the host galaxy and cluster environment by at least two orders of magnitude, even assuming perfectly efficient energy transfer acting over the required scales.  In addition, the observation of significant cooling in IC-scattering of CMB photons is evidence that a significant fraction of this energy does not get injected into the local environment.  

Nevertheless, IC scattering could provide a very efficient mechanism for producing hard, ionising photons on scales far from the central AGN.  Given reasonable assumptions for the halo gas density, X-ray luminosity at the level observed in 4C 23.56 could potentially ionize a large volume of halo gas over scales of several hundred kpc, as outlined in Scharf et al. (2003).  Moreover, several additional mechanisms could increase the feedback potential of the X-ray lobes dramatically.  Energy stored within the magnetic field, found to be of order $E_{e}$ in the Erlund et al. (2006) sources, may increase the total energy of the jet.  More significant may be the $pdV$ work associated with the expansion of lobes into the surrounding medium, which has been estimated to provide energies of up to $10^{61}$ erg in lower redshift radio galaxies and observed to significantly heat the surrounding intracluster medium (e.g., B{\^i}rzan et al., 2004; McNamara et al., 2005).  Finally, bursts of AGN activity may occur several times over the course of a Gyr, re-accelerating the radio lobes and driving subsequent periods of environmental feedback.

\section*{Acknowledgments}
OJ acknowledges PPARC for support through a Postdoctoral Fellowship.  OA and PNB thank the Royal Society for support through its University Research Fellowship Scheme.

\end{document}